\documentclass[lineno]{jfm}
\usepackage{graphicx}
\usepackage{mathtools}
\usepackage{labelfig}
\usepackage[svgnames,x11names,table,dvipsnames]{xcolor}
\usepackage{epstopdf}

\usepackage{natbib}
\usepackage[justification=justified]{caption}
\usepackage{amsmath,bm}
\usepackage{mathrsfs} 
\usepackage{rotating}

\usepackage[colorlinks=true,linkcolor=blue,citecolor=blue,urlcolor = blue]{hyperref}
\usepackage[nameinlink]{cleveref}
\crefname{equation}{equation}{equation}
\crefname{Equation}{Equation}{Equation}
\crefname{eqnarray}{}{}
\crefname{figure}{figure}{figures}
\Crefname{Figure}{Figure}{Figures}
\crefname{table}{table}{tables}

\usepackage{amsfonts}
\usepackage{booktabs}
\usepackage{lipsum}

\usepackage{color}
\usepackage{soul} 

\definecolor{darkgreen}{rgb}{0,0.5,0.0}
\definecolor{purple}{rgb}{0.7,0.0,0.5}

\newcommand{\poscite}[1]{\citeauthor{#1}'s (\citeyear{#1})}

\newcommand{\RomanNumeralCaps}[1]

\title{Energetics of particle-size segregation}

\author{Tom\'as Trewhela\aff{1}\corresp{\email{tomas.trewhela@uai.cl}}
\and
Hugo N. Ulloa\aff{2}\corresp{\email{ulloa@sas.upenn.edu}}}
\affiliation{
\aff{1} Facultad de Ingenier\'ia y Ciencias, Universidad Adolfo Iba{\~n}ez, Vi{\~n}a del Mar, Chile.

\aff{2} Department of Earth and Environmental Science, University of Pennsylvania, Philadelphia, USA.}

\begin{document}
\maketitle

\begin{abstract}
We introduce a continuum framework for the energetics of particle-size segregation in bidisperse granular flows. Building on continuum segregation equations and a recent segregation flux model, the proposed framework offers general analytical expressions to study the physics of granular flows from a mechanical energy perspective. To demonstrate the framework's applicability, we examined the energetics in shear-driven flows. Numerical experiments with varying frictional coefficients and particle-size ratios revealed two distinct phases in the associated energetics with particle segregation and diffusive remixing, and that the potential energy to the kinetic energy ratio in the steady state follows the scaling relationship $\hat{E}^{(s)}_{gp} / \hat{E}^{(s)}_{k} \propto Pe^{-1/2}_{sr}$ for $0.4 \leq Pe_{sr} \leq 300$, the segregation-rheology Péclet number. Our findings hint that the bulk segregation-mixing state can be predicted and controlled using $Pe_{sr}$, determined from known system parameters, providing a impactful tool for engineering and geophysical applications.
\end{abstract}

\vspace{-1.5cm}
\section{Introduction} 
\label{sec:Introduction}
Granular materials segregate and mix as they flow \citep{Gray18,Umbanhowar19}. Although no general agreement exists on the determining drivers, relative motion between particles of differing size conforming a granular bulk is essential for segregation, and they occur when available mechanical energy—sourced from external forcing or gravity—is transformed into kinetic energy. However, how the available energy partitions between segregation, mixing and interparticle friction within the granular bulk? Addressing this question is crucial not only for  understanding of granular flow dynamics but also for optimising engineering processes aiming to transport or mix polydisperse materials.

Research on polydisperse granular flows from a mechanical energy perspective remains limited. Most of the studies have focused on characterising the energy dissipated by friction \cite[e.g.][]{pereira2017,jiang2018}, yet only a few have connected the energy with granular flow regimes. Using discrete element simulations, \cite{sun2013} found that the effective friction coefficient in sheared bidisperse granular flows is inversely proportional to the ratio of elastic to kinetic energy, and emphasised that energetics studies could bridge the mechanics from quasi-static to rapid flow regimes. Despite these insights, a comprehensive understanding of how mechanical energy transforms and relates to segregation-mixing states in flowing granular mixtures is still lacking.

Combining segregation theory \citep{Gray18} with laboratory experiments, \poscite{Trewhela21a} scaling law for particle-size segregation provides robust predictions for segregation in dense granular flows from various physical parameters. Yet, advancing the modelling of granular flows in complex configurations requires characterising the energy source, how this energy partitions and transforms within and by the granular bulk. Having an energetics framework for granular flows is, therefore, instrumental to quantify how much of the available energy in a system leads to particles' motion, segregation, mixing and irreversible dissipation due to friction.

This paper introduces a new continuum framework for characterising particle-size segregation and mixing from an energetics perspective. In \S~\ref{sec:theory}, we present the convection-diffusion model for particle-size segregation based on continuum-mixture theory \citep{Gray18}. Building on this and a recent segregation scaling law \citep{Trewhela21a}, we derive the energetics for noncohesive, inelastic, bidisperse particle flows in \S~\ref{sec:3}. In \S~\ref{sec:4}, we illustrate the framework's applicability by studying the energetics for shear-driven, bidisperse granular flows, revealing a new scaling relationship between the ratio of the mechanical energy components and the segregation-rheology Péclet number, which offers potential applications in industrial and geophysical processes.
\vspace{-.5cm}
\section{Theoretical framework}
\label{sec:theory}

\subsection{Bidisperse particle-size segregation equations}
\label{sec:bpseg}

We consider a granular medium of bidisperse spherical particles\textemdash of differing small $d_{s}$ and large $d_{l}>d_{s}$ diametres but uniform density $\rho_{*}$\textemdash within a fixed volume $V$. The mass of the mixture is $M = M_{l}+M_{s}$, with $M_{s} = \rho_{*}n_{s} (\pi/6)d^{3}_{s}$ and $M_{l} = \rho_{*}n_{l} (\pi/6)d^{3}_{l}$ the mass of the small particles and the large particles, respectively. The solids volume fraction $\Phi$ is then determined by dividing $M$ by the reference mass $M_{*}=\rho_{*}\,V$, so $\Phi=\Phi_{l}+\Phi_{s}$. In dynamic systems, each concentration $\Phi_{\nu}$ ($\nu\in\left\{s,l\right\}$) is, in principle, space- and time-dependent. However, the concentration $\Phi$ usually varies slightly in most dense granular flows. By assuming $\Phi$ to be constant, we can express each particle size concentration in terms of its corresponding species concentration, $\phi_{s}=\Phi_{s}/\Phi$ and $\phi_{l}=\Phi_{l}/\Phi$. Thus, $\phi_{s}+\phi_{l}=1$ establishes the inherent mass conservation law for the granular system. 

The bidisperse granular material can be treated as a continuum medium, for which mixture theory applied to the granular bulk yields the convective-diffusive equations for segregation,
\begin{subeqnarray}
    \frac{\partial \phi_{s}}{\partial t}+\nabla\cdot\left(\phi_{s}\bm{u}\right)+\nabla\cdot \left(f_{sl}\phi_{s}\phi_{l}\frac{\bm{g}}{|\bm{g}|}\right)&=&\nabla\cdot(\mathcal{D}_{sl}\nabla\phi_{s}),\label{eq:phis}\\
    \frac{\partial \phi_{l}}{\partial t}+\nabla\cdot\left(\phi_{l}\bm{u}\right)-\nabla\cdot \left(f_{sl}\phi_{s}\phi_{l}\frac{\bm{g}}{|\bm{g}|}\right)&=&\nabla\cdot(\mathcal{D}_{sl}\nabla\phi_{l}),\label{eq:phil}
\end{subeqnarray}
where $\bm{u}$ is the divergence-free granular bulk velocity, $\mathcal{D}_{sl}$ is the particle diffusivity and 
$f_{sl}\phi_{s}\phi_{l}\,\bm{g}/|\bm{g}|$ characterises the segregation flux, with $f_{sl}$ the particle-size segregation velocity magnitude \citep{Gray18}. The gravitational norm vector $\bm{g}/|\bm{g}|$ is included to confer a direction to size segregation, determined by the gravity-driven kinetic sieving and squeeze expulsion mechanisms. The functional form of $f_{sl}$ has been proposed to be quadratic or cubic, symmetrical or asymmetrical \citep{Gajjar14,Umbanhowar19}. Recent scaling laws have aimed to parameterise $f_{sl}$ in terms of physical quantities, such as pressure $p$, shear rate $\dot{\gamma}$, local particle concentrations $\phi_{s}$ and particle diameter $d$. With that goal, \citet{Trewhela21a} proposed that the segregation velocity $f_{sl}$ could be described by
\begin{equation}
f_{sl}=\left(\frac{\mathcal{B}\,\rho_{*}\,g\,\dot{\gamma}\,\bar{d}^{2}}{p}\right)\left[(R_{d}-1)+\mathcal{E}(1-\phi_{s})(R_{d}-1)^{2}\right],
    \label{eq:fsl}
\end{equation}
where $\mathcal{B}$ and $\mathcal{E}$ are empirical constants, $\bar{d}=\phi_{s}d_{s}+\phi_{l}d_{l}$ denotes the particles' concentration-averaged diameter, whereas $R_{d}=d_{l}/d_{s}$ defines the particles' size ratio. Merging the latter two definitions, the concentration-averaged diameter can be expressed as $\bar{d}=(1-\chi_{d}\,\phi_{s})R_{d}\,d_{s}$, which captures the asymmetric behaviour of size segregation by defining $\chi_{d}=(R_{d}-1)/R_{d}$ as the asymmetry coefficient \citep{Gajjar14,vanderVaart15,Trewhela24}. Note that the function \eqref{eq:fsl} is not particular or case-dependent; it describes the segregation velocity for both small and large particles, at variable concentrations $\phi_{s}$ and $\phi_{l}=(1-\phi_{s})$, respectively. It also compiles a comprehensive set of experimental and numerical observations hinting pressure $p$, shear rate $\dot{\gamma}$ and size ratio $R_{d}$ dependence \citep{Golick09,Thornton12,Fry18,Trewhela21b}.

From the definition of $\Phi_{\nu}$, \eqref{eq:phis} can be directly expressed as
\begin{equation}
    \frac{\partial \Phi_{\nu}}{\partial t}+\nabla\cdot\left(\Phi_{\nu}\,\bm{u}\right)+\nabla\cdot \bm{F}_{\Phi_{\nu}}=\nabla\cdot(\mathcal{D}_{sl}\nabla\Phi_{\nu}),
    \label{eq:PhiSmall}
\end{equation}
where the segregation flux $\bm{F}_{\Phi_{\nu}}$ for the $\nu$-particle species is defined as
\begin{equation}
    \bm{F}_{\Phi_{\nu}} = \begin{cases}
+f_{sl}\,\Phi_{s}\,[1-\phi_{s}]\frac{\bm{g}}{|\bm{g}|} &\textrm{for}\,\,\nu=s,\\
-f_{sl}\,\Phi_{l}\,[1-\phi_{l}]\frac{\bm{g}}{|\bm{g}|} &\textrm{for}\,\, \nu=l,
\end{cases}
\end{equation}
whereas the particles' diffusivity $\mathcal{D}_{sl}=\mathcal{A}\dot{\gamma}\bar{d}^{2}$ \citep{Utter04} is responsible for diffusive remixing and controls the final segregation state. 
\vspace{-0.2cm}

\section{Energetics for the $\nu$-particle size segregation}\label{sec:3}

Assuming rigid inelastic grains, the potential energy for the $\nu$-particle species in the volume $V$ of domain $\Omega$ is only determined by the gravitational component,
\begin{equation}\label{eq:GPE}
    E^{(\nu)}_{gp} = \int_{\Omega} \rho_{*}\,g\,z\,\Phi_{\nu}\,{\rm d}V,
\end{equation}
with $\mathscr{E}^{(\nu)}_{gp}(t,\bm{x}) = \rho_{*}\,g\,z\,\Phi_{\nu}(t,\bm{x})$ the gravitational potential energy (GPE) density and $z$ the vertical coordinate pointing upwards. Likewise, we define the kinetic energy (KE) of the $\nu$-particle species  within the volume $V$ as
\begin{equation}\label{eq:GPE}
    E^{(\nu)}_{k} = \frac{1}{2}\int_{\Omega} \rho_{*}\,\Phi_{\nu}\, |\bm{u}|^{2} {\rm d}V,
\end{equation}
with $\mathscr{E}^{(\nu)}_{k}(t,\bm{x}) = \frac{1}{2}\rho_{*}\Phi_{\nu}(t,\bm{x})\,|\bm{u}(t,\bm{x})|^{2}$ the KE density. 

\subsection{Energetics for the gravitational potential energy} \label{sec:3.1}
From \eqref{eq:PhiSmall}, the evolution equation for the GPE density of the $\nu$-particle species is given by 
\begin{equation}
    \frac{\partial \mathscr{E}^{(\nu)}_{gp}}{\partial t} = \rho_{*}\,g\,z\left[- \nabla\cdot\left(\Phi_{\nu}\,\bm{u}_{\nu}\right)-\nabla\cdot\bm{F}_{\Phi_{\nu}}+\nabla\cdot\left(\mathcal{D}_{sl}\nabla\Phi_{\nu}\right)\right].
\end{equation}
Therefore, utilising vector identities and the divergence theorem, the evolution equation for the GPE in the volume $V$ unfolds as follows: 
\begin{equation}\label{eq:Egp}
\begin{split}
        \frac{{\rm d}E^{(\nu)}_{gp}}{{\rm d}t} =& \underbrace{-\oint_{\partial\Omega}\rho_{*}\,g\,z\,\left[\Phi_{\nu}\,\bm{u}+\bm{F}_{\Phi_{\nu}}-\mathcal{D}_{sl}\nabla\Phi_{\nu}\right]\cdot\hat{\bm{n}}\,{\rm d}S}_{\Sigma^{(\nu)}_{gp}}\\
        & +\underbrace{\int_{\Omega}\rho_{*}\,g\,\Phi_{\nu}\,w\,{\rm d}V}_{\Psi^{(\nu)}_{c}} +\underbrace{\int_{\Omega}\rho_{*}\,g\,\bm{F}_{\Phi_{\nu}}\cdot\hat{\bm{k}}\,{\rm d}V}_{\Psi^{(\nu)}_{gps}} -\underbrace{\int_{\Omega}\rho_{*}\,g\,\mathcal{D}_{sl}\frac{\partial\Phi_{\nu}}{\partial z}\,{\rm d}V}_{\Psi^{(\nu)}_{gpd}}.
\end{split}
\end{equation}
Thus, the rate of change of $E^{(\nu)}_{gp}$ is determined by four main energetics, which characterise the injection, transformation and dissipation of GPE. The first energetics in the right-hand side of \eqref{eq:Egp} characterises the net surface boundary flux $\Sigma^{(\nu)}_{gp}$ that integrates fluxes resulting from advection, segregation and diffusion. The second energetics $\Psi^{(\nu)}_{c}$ quantifies the rate of change of $E^{(\nu)}_{gp}$ due to particle redistribution resulting from vertical convective flows. Whereas the third $\Psi^{(\nu)}_{s}$ and fourth $\Psi^{(\nu)}_{gpd}$ energetics characterise the rate of change of $E^{(\nu)}_{gp}$ owing to vertical particle segregation and diffusion, respectively. 
\vspace{-0.3cm}

\subsection{Energetics for the kinetic energy} \label{sec:3.2}
The evolution equation for the KE density is derived from 
\begin{equation}
    \frac{\partial \mathscr{E}^{(\nu)}_{k}}{\partial t} = \rho_{*} \frac{1}{2}|\bm{u}|^{2}\,\frac{\partial \Phi_{\nu}}{\partial t} +\rho_{*}\,\Phi_{\nu}\,\bm{u}\cdot \frac{\partial \bm{u}}{\partial t}
\end{equation}
and the linear momentum equation for the particles mixture,
\begin{equation}\label{eq:LM}
    \rho_{*}\Phi\left[\frac{\partial \bm{u}}{\partial t} + \nabla\left(\frac{1}{2}|\bm{u}|^{2}\right) \right] = \nabla\cdot\mathsfbi{T}+\rho_{*}\Phi\,\bm{g}.
\end{equation}
The stress tensor can be decomposed into $\mathsfbi{T} = \bm{\tau}-p\mathsfbi{1}$ with $p$ the pressure multiplying the tensor identity and $\bm{\tau}$ the deviatoric shear stress. Consequently, from \eqref{eq:PhiSmall} and \eqref{eq:LM}, the evolution equation of $\mathscr{E}^{(\nu)}_{k}(t,\bm{x})$ is given by
\begin{equation}\label{eq:dEk}
\begin{split}
\frac{\partial \mathscr{E}^{(\nu)}_{k}}{\partial t} = & \frac{1}{2}|\bm{u}|^{2}\,\rho_{*}\left\{ - \nabla\cdot\left(\Phi_{\nu}\,\bm{u}\right)-\nabla\cdot\bm{F}_{\Phi_{sl}}+\nabla\cdot\left(\mathcal{D}_{sl}\nabla\Phi_{\nu}\right)\right\}\\
& + \bm{u}\cdot\left\{-\rho_{*}\Phi_{\nu}\,\nabla\left(\frac{1}{2}|\bm{u}|^{2}\right)+\phi_{\nu}\nabla\cdot\mathsfbi{T} +\rho_{*}\,\Phi_{\nu}\,\bm{g}\right\}.
\end{split}
\end{equation}
Integrating \eqref{eq:dEk} in the volume $V$, the evolution equation for the KE is given by
\begin{equation}\label{eq:dEkdt}
\begin{split}
\frac{{\rm d}E^{(\nu)}_{k}}{{\rm d}t} =&
\underbrace{-\int_{\partial\Omega}\left[\frac{\rho_{*}}{2}|\bm{u}|^{2}\left(\Phi_{\nu}\,\bm{u}+\bm{F}_{\Phi_{\nu}}-\mathcal{D}_{sl}\nabla\Phi_{\nu}\right)-\left(\phi_{\nu}\,\bm{u}\cdot\mathsfbi{T}\right)\right]\cdot\hat{\bm{n}}\,{\rm d}S}_{\Sigma^{(\nu)}_{k}}\\
 &
+\underbrace{\int_{\Omega}\bm{F}_{\Phi_{sl}}\cdot\nabla\left(\frac{\rho_{*}}{2}|\bm{u}|^{2}\right){\rm d}V}_{\Psi^{(\nu)}_{ks}}- \underbrace{\int_{\Omega}\mathcal{D}_{sl}\nabla\Phi_{\nu}\cdot\nabla\left(\frac{\rho_{*}}{2}|\bm{u}|^{2}\right){\rm d}V}_{\Psi^{(\nu)}_{kd}}\\
& -\Psi^{(\nu)}_{c} -\underbrace{\int_{\Omega}\nabla\left(\phi_{\nu}\,\bm{u}\right):\mathsfbi{T}\,{\rm d}V}_{\Psi^{(\nu)}_{\mu}}.
\end{split}
\end{equation}
The rate of change of KE due to boundary fluxes is determined by the expression $\Sigma^{(\nu)}_{k}$, while changes due to segregation, diffusion, and vertical transport are determined by the expressions $\Psi^{(\nu)}_{ks}$, $\Psi^{(\nu)}_{kd}$, and $\Psi^{(\nu)}_{c}$, respectively. Lastly, the dissipation rate of KE due to friction is determined by the expression $\Psi^{(\nu)}_{\mu}$.
\vspace{-0.2cm}

\section{Application: shear-driven granular flow}\label{sec:4}
The framework introduced in \S~\ref{sec:3} allows investigating granular flows from an energy perspective. We illustrate its applicability by examining the energetics of sheared granular flows undergoing segregation and mixing dynamics. For such configuration, we must have a constitutive law relating stresses and the granular mixture rheology to determine the energy injection and dissipation resulting from particles' friction, $\Psi^{(\nu)}_{\mu}$; which are defined next.

\subsection{Stress tensor and $\mu(I)$-rheology for bidisperse granular mixtures}
\label{sec:stresstensor}
To calculate $\Sigma^{(\nu)}_{k}$ and $\Psi^{(\nu)}_\mu$ in \eqref{eq:dEkdt}, the stress tensor $\mathsfbi{T}$ needs to be described thoroughly. A first relationship to do so is the $\mu(I)$-rheology, that relates the shear $\tau$ and normal $p$ stresses through a Coulombic $\tau=\mu(I)\,p$. Embedded within this definition, the frictional coefficient $\mu(I)$ is found to be dependent on the inertial number $I=\dot{\gamma}\,\bar{d}/\sqrt{p/\rho^{*}}$ \citep[e.g.][]{GDRMidi04}. The frictional coefficient is obtained by fitting the empirically-determined function $\mu(I) = \mu_{1} +(\mu_{2}-\mu_{1})/(I_{0}/I+1)$, where $\mu_{1}$ and $\mu_{2}$ are the static and dynamic frictional coefficients, respectively \citep{Jop06}. These two coefficients alongside the material parameter $I_{0}$ correspond to the empirical parameters that define the $\mu(I)$-rheology. However, the above relationship for $\mu(I)$ is ill posed. \cite{Barker17} partially regularised the $\mu(I)$ definition with the extended relationship
\begin{equation}
    \mu(I)=\begin{dcases}
    \sqrt{\alpha/\text{log}\left(A/I\right)}, & \text{ for } I\leq I_{1}\\
          \frac{\mu_{1}I_{0}+\mu_{2}I+\mu_{\infty}I^{2}}{I+I_{0}}, & \text{ for } I> I_{1},
        \end{dcases}
 \label{eq:muIreg}
\end{equation}
where $\mu_{\infty}$, $\alpha$ and $I_{1}$ are material-dependent constants. The two latter parameters with $A$ serve as fitting constants for the partially-regularised $\mu(I)$-rheology which guarantee continuity of \eqref{eq:muIreg}. For this work, the values for all these parameters and constants defining the $\mu(I)$-rheology are presented in table \ref{tab:params}.

\begin{table}
\captionsetup{width=1\linewidth}
  \begin{center}
  $\mu_{1}=0.342$,\enspace $\mu_{2}=0.557$,\enspace $I_{0}=0.069$, \enspace $\alpha=1.9$,\enspace$\mu_{\infty}=0.05$,\enspace $I_{1}=0.005$,\enspace $A=1.5647\times10^{4}$\\
  $d_{l}=143$ $\rm \mu m$,\enspace $\rho_{*}=2500$ kg m$^{-3}$,\enspace $\Phi=0.6$,\enspace  $p_{0}=300$ Pa, \enspace $u_{0}=0.2$ m s$^{-1}$, \enspace $\chi_{\mu}=0.0075$\\
  $\mathcal{A}=0.108$,\enspace $\mathcal{B}=0.374$,\enspace $\mathcal{E}=2.096$,

  \caption{Parameters used in this work. Frictional parameters  $\mu_{1}$, $\mu_{2}$ and $I_{0}$ for the $\mu(I)$-rheology \citep{Jop06}. Coefficient $\mu_{\infty}$ with the fitting parameters $\alpha$, $I_{1}$ and $A$ for the partial regularisation of the $\mu(I)$-rheology were measured by \cite{Barker17} for $d_{l}=$143 µm glass beads. Grains' intrinsic density $\rho_{*}$; solids volume fraction $\Phi$; the specific values for pressure $p_{0}$ and velocity $u_{0}$ at the wall; and, the frictional asymmetry coefficient $\chi_{\mu}$ \citep{Trewhela24}. Finally the constants for  diffusivity $\mathcal{A}$ \citep{Utter04} and segregation $\mathcal{B}$, $\mathcal{E}$ \citep{Trewhela21a}.}
  \label{tab:params}
  \end{center}
\end{table}

To include bidispersity effects in the bulk's frictional response, we define a concentration-averaged frictional coefficient $\bar{\mu}=\mu_{s}\phi_{s}+\mu_{l}\phi_{l}=(1-\chi_{\mu}\phi_{s})R_{\mu}\mu_{s}$. In the latter, $\mu_{\nu}$ denotes the frictional coefficient for the $\nu$ species and $R_{\mu}=\mu_{l}/\mu_{s}$ a ratio that defines the asymmetry parameter, $\chi_{\mu}=(R_{\mu}-1)/R_{\mu}$ \citep{Trewhela24} (table \ref{tab:params}). 
From the above definitions, the stress tensor components are given by
\begin{equation}
T_{ij}=\begin{dcases}
    \bar{\mu}(I)p-p & \text{ for }i=j\\
          \bar{\mu}(I)p, & \text{ for } i\neq j,
        \end{dcases}, \hspace{0.5cm}{\rm with}\hspace{0.5cm} i,j=\{x,z\},
\end{equation}
which is naturally averaged over the concentrations of particle species.
\vspace{-0.2cm}

\subsection{Energetics of a sheared bidisperse granular layer}
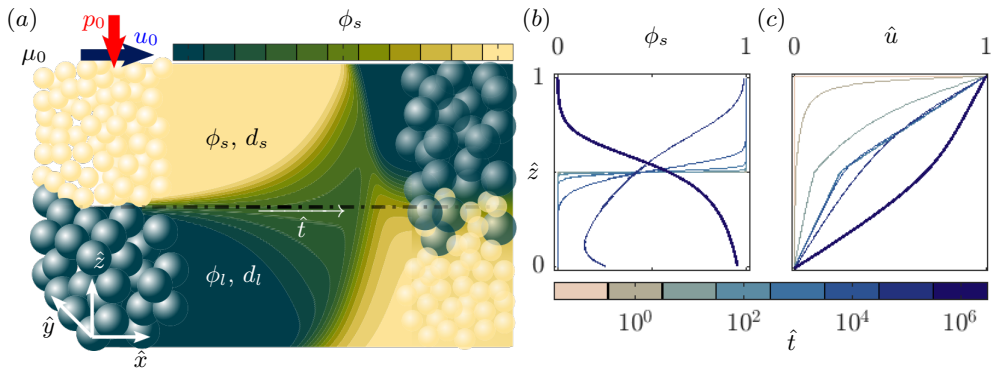
\begin{figure}
\raggedright
\captionsetup{width=\linewidth}
\SetLabels
\L (0.005*0.95) $(a)$\\
\L (0.51*0.95) $(b)$\\
\L (0.74*0.95) $(c)$\\
\L (0.20*0.20) $\textcolor{white}{\phi_{l},\,d_{l}}$\\
\L (0.20*0.6) $\textcolor{black}{\phi_{s},\,d_{s}}$\\
\L (0.77*0.0) $\hat{t}$\\
\L (0.13*-0.05) $\hat{x}$\\
\L (0.04*0.05) $\hat{y}$\\
\L (0.09*0.24) $\textcolor{white}{\hat{z}}$\\
\L (0.08*0.95) $\textcolor{red}{p_{0}}$\\
\L (0.13*0.9) $\textcolor{blue}{u_{0}}$\\
\L (0.02*0.85) $\textcolor{black}{\mu_{0}}$\\
\L (0.25*0.39) $\textcolor{white}{\xrightarrow{\quad \quad \quad}}$\\
\L (0.29*0.33) $\textcolor{white}{\hat{t}}$\\
\L (0.515*0.5) $\hat{z}$\\
\L (0.33*0.95) $\phi_{s}$\\
\L (0.63*0.90) $\phi_{s}$\\
\L (0.865*0.90) $\hat{u}$\\
\endSetLabels
\strut\AffixLabels{\centerline{\includegraphics[height=4.5cm]{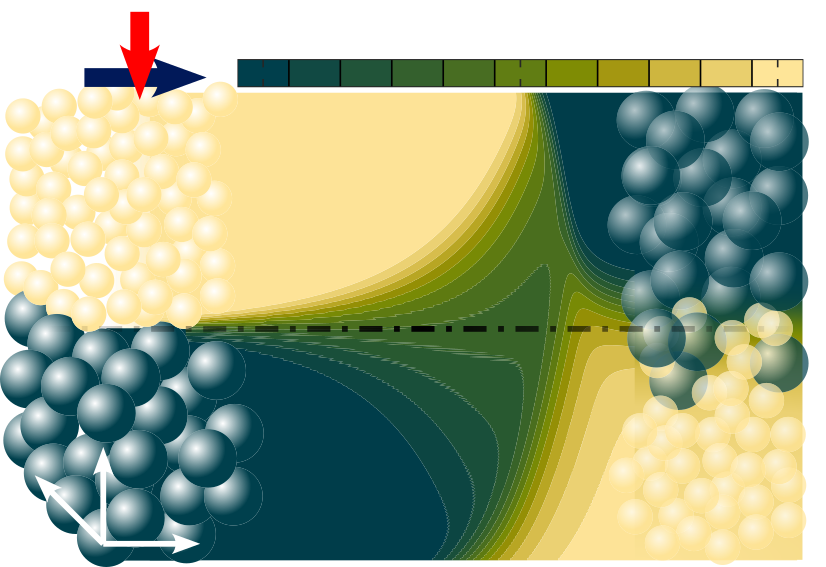}\includegraphics[height=4.5cm]{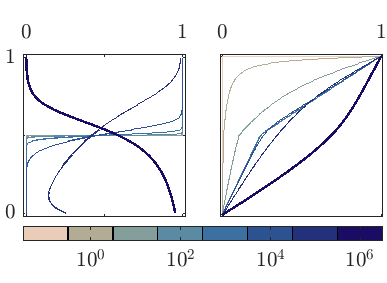}}}
\caption{(\emph{a}) Schematic and numerical results. (\emph{b-c}) Temporal evolution of the small particle species concentration $\phi_{s}(\hat{t},\hat{z})$ and velocity profile $\hat{u}(\hat{t},\hat{z})$ for $R_{d}=2.5$, $\mu_{0}=0.5$.}
\label{fig:numsol}
\end{figure}

As case study, we consider a bidisperse granular bulk of depth $h$ sustaining shear exerted by a top plate imposing a uniform normal pressure distribution $p=p_{0}$ and moving at a constant velocity $u_{0}$ in the streamwise $x$-direction (\Cref{fig:numsol}\emph{a}). Based on the rheological model in~\S~\ref{sec:stresstensor}, the pressure, velocity and frictional coefficient $\mu_{0}$ set a shear $\tau_{xz} = \tau_{0}=p_{0}\,\mu_{0}$ (table \ref{tab:params}) at the top boundary. In contrast, the granular material in contact with the motionless bottom does not slip nor roll. Under these conditions, the granular system develops a zero-gravity shear flow, which, for simplicity, is treated as two-dimensional and periodic in the $x$-direction. As a result, the sheared layer depends on the upper boundary condition:  $h=(\mu_{0}/\mu_{1}-1)p_{0}/(\rho_{*}\,g\,\Phi)$ \citep{Jerome18}. The top and bottom boundaries are closed, so fluxes across them are null. Without loss of generality, we examine the momentum-segregation equations of the small particle species, i.e. $\nu=s$, which in nondimensional form are given by
\begin{equation}
\frac{\partial \hat{u}}{\partial \hat{t}}=\frac{\partial}{\partial \hat{z}}\left(\left[1-\chi_{\mu}\,\phi_{s}\right]R_{\mu}\,\mu_{s}\, \hat{p}\right), \quad \frac{\partial \phi_{s}}{\partial \hat{t}}=\frac{\partial}{\partial \hat{z}}\left(\hat{f}_{sl}\,\phi_{s}[1-\phi_{s}]
+\hat{\mathcal{D}}_{sl}\frac{\partial\phi_{s}}{\partial \hat{z}}\right),
\label{eq:nseg}
\end{equation}
where $\hat{u}=u/u_{0}$, $\hat{p}=p/p_{0}$, $\hat{z}=z/h$, $\hat{t}=t/(h/u_{0})$, $\hat{f}_{sl}=f_{sl}/u_{0}$, and $\hat{\mathcal{D}}_{sl}=\mathcal{D}_{sl}/(u_{0}\,h)$ denote the nondimensional variables \citep{Trewhela24}. 
From \eqref{eq:nseg} and considering the energy scale per unit area, `$\Phi\,\rho_{*}\,g\,h^{2}$', the evolution equation for the nondimensional GPE, $\hat{E}^{(s)}_{gp}=\int^{1}_{0}\phi_{s}\hat{z}\,{\rm d}\hat{z}$, reduces to 
\begin{equation}\label{eq:n_dgpe_dt}
   \frac{{\rm d}{\hat{E}^{(s)}}_{gp}}{{\rm d}t} = \underbrace{\int^{1}_{0}-\hat{f}_{sl}\,\phi_{s}\left[1-\phi_{s}\right]{\rm d}\hat{z}}_{\hat{\Psi}^{(s)}_{gps}}-\underbrace{\int^{1}_{0}\hat{\mathcal{D}}_{sl}\frac{\partial\phi_{s}}{\partial \hat{z}}\,{\rm d}\hat{z}}_{\hat{\Psi}^{(s)}_{gpd}}.    
\end{equation}
Likewise, considering the same energy scale, the evolution equation for the nondimensional KE, $\hat{E}^{(s)}_{ke}=\frac{1}{2}Fr^{2}\int^{1}_{0}\hat{u}^{2}\,{\rm d}\hat{z}$, simplifies to
\begin{equation}\label{eq:n_dke_dt}
    \frac{{\rm d}\hat{E}^{(s)}_{k}}{{\rm d}\hat{t}}=\underbrace{Fr^{2}\left(\hat{u}\left[1-\chi_{\mu}\phi_{s}\right]R_{\mu}\,\mu_{s}\,\hat{p}\right)\big|_{\hat{z}=1}}_{\hat{\Sigma}^{(s)}_{k}} - \underbrace{Fr^{2}\int^{1}_{0}\left[1-\chi_{\mu}\phi_{s}\right]R_{\mu}\,\mu_{s}\,\hat{p}\,\frac{\partial \hat{u}}{\partial \hat{z}}\,{\rm d}\hat{z}}_{\hat{\Psi^{(s)}_{\mu}}},
\end{equation}
with $Fr = u_{0}/\sqrt{g\,h}$ the Froude number. The energetics $\hat{\Sigma}^{(s)}_{k}$ and $\Psi^{(s)}_{\mu}$ are positive definite and determine the rate of kinetic (mechanical) energy production due to surface shear and KE dissipation owing to friction, respectively. In contrast, the energetics $\hat{\Psi}^{(s)}_{gps}$ and $\hat{\Psi}^{(s)}_{gpd}$ quantify the transformation rate of potential (mechanical) energy due to vertical segregation and diffusion of particles, respectively; their signs are not fixed. The energy pathways controlled by the processes responsible for transforming the mechanical energy\textemdash i.e. segregation, particle diffusion and friction\textemdash are unknown and investigated next.
\subsection{Numerical experiments}
Following \cite{Trewhela24}, we resolved numerically the coupled, nonlinear PDE system in \eqref{eq:nseg} using the Method of Lines (MOL). This numerical scheme is unconditionally-stable and considered a constant pressure distribution $p=p_{0}$ so its robustness is not compromised by the ill-posed $\mu(I)$-rheology, despite it is regularised in \eqref{eq:muIreg}. A total of 18 simulations were ran, considering first a set of parameters $\mu_{0}\times R_{d}=\left\{0.5, 0.8\right\}\times\left\{1.25,1.5,1.75,2.0,2.25,2.5\right\}$ yielding 12 simulations to understand midrange values and secondly the set $\left\{0.4, 1.2, 1.6\right\}\times\left\{1.25,2.5\right\}$ to explore extreme values for $\mu_{0}$. The resulting set sheds light on the energetics under extreme frictional regimes and a wide range of segregation-mixing dynamics. Here, the fundamental parameter is the segregation-rheology P\'eclet number \citep{Trewhela24}, which spans as $Pe_{sr}=Pe/Sc\in\left[0.4,300\right]$, with $Pe=f_{sl}\,h/\mathcal{D}_{sl}$ the P\'eclet number and $Sc=\nu_{g}/\mathcal{D}_{sl}$ the Schmidt number. \Cref{fig:numsol} illustrates an example of the spatiotemporal evolution of small-particle concentration $\phi_{s}(\hat{t},\hat{z})$ and vertical profiles of the shear-driven granular flow $\hat{u}(\hat{t},\hat{z})$, which are utilised to compute the energetics in \eqref{eq:n_dgpe_dt} and \eqref{eq:n_dke_dt}. 
\begin{figure}
\raggedright
\captionsetup{width=\linewidth}
\SetLabels
\L (0.010*0.98) $(a)$\\
\L (0.015*0.15) \rotatebox{90}{$\rm GPE~energetics\times10^{5}$}\\
\L (0.105*0.45) 
$\textcolor{BrickRed}{\hat{\Psi}^{(s)}_{gps}}$\\
\L (0.20*0.45) 
$\textcolor{blue}{\hat{\Psi}^{(s)}_{gpd}}$\\
\L (0.195*-0.02) $\hat{t}$\\
\L (0.135*0.275) $\rightarrow$\\
\L (0.190*0.275) $\leftarrow$\\
\L (0.165*0.195) $R_{d}$\\
\L (0.195*-0.02) $\hat{t}$\\
\L (0.34*0.98) $(b)$\\
\L (0.345*0.29) \rotatebox{90}{$\rm KE~energetics$}\\
\L (0.51*-0.02) $\hat{t}$\\
\L (0.57*0.45) $\textcolor{OliveGreen}{\hat{\Sigma}^{(s)}_{k}}$\\
\L (0.465*0.45) $-\textcolor{Mulberry}{\hat{\Psi}^{(s)}_{\mu}}$\\
\L (0.50*0.275) $\rightarrow$\\
\L (0.56*0.275) $\leftarrow$\\
\L (0.53*0.195) $R_{d}$\\
\L (0.65*0.98) $(c)$\\
\L (0.75*0.68) \rotatebox{0}{$\frac{{\rm d}\hat{E}^{(s)}_{k}}{{\rm d}\hat{t}}$}\\
\L (0.72*0.25) \rotatebox{0}{$\frac{{\rm d}\hat{E}^{(s)}_{gp}}{{\rm d}\hat{t}}\times 10^{5}$}\\
\L (0.83*-0.02) $\hat{t}$\\
\L (0.91*0.75) $R_{d}$\\
\endSetLabels
\strut\AffixLabels{\centerline{\includegraphics[width=0.99\linewidth]{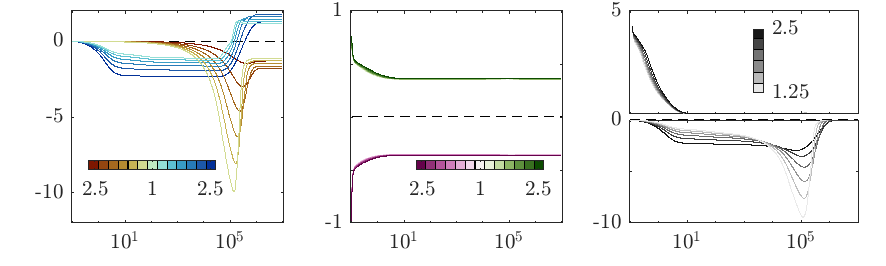}}}
\caption{Numerical results for $\mu_{0}\times R_{d}=\left\{0.8\right\}\times\left\{1.25,1.5,1.75,2.0,2.25,2.5\right\}$. (\emph{a}) Energetics of GPE, $\hat{\Psi}^{(s)}_{gps}$ and $\hat{\Psi}^{(s)}_{gpd}$ vs time $\hat{t}$. (\emph{b}) Energetics of KE, $\hat{\Sigma}^{(s)}_{k}$ and $\hat{\Psi}^{(s)}_{\mu}$vs time. (\emph{c})  KE budget (top panel) and GPE budget (bottom panel) vs time $\hat{t}$.}
\label{fig:energetics}
\end{figure}
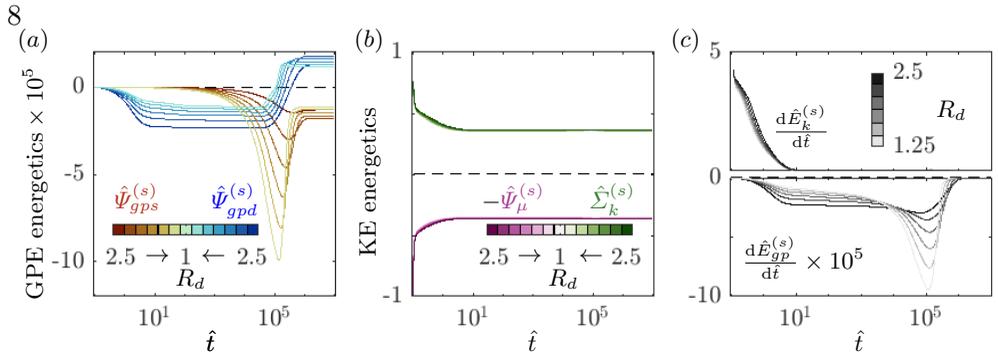

\Cref{fig:energetics}($a$) shows time series of the segregation energy flux $\hat{\Psi}^{(s)}_{gps}$ and the diffusive energy flux $\hat{\Psi}^{(s)}_{k}$ for $\mu_{0}=0.8$ and $1.25\leq R_{d}\leq 2.5$. The energy flux $\hat{\Psi}^{(s)}_{gps}$ is always negative, i.e. segregation continuously consumes the available GPE that the small particles have. Yet, $\hat{\Psi}^{(s)}_{gps}$ exhibits two striking phases. The first phase is characterised by an exponential-like increase of the segregation flux, supported by the diffusive flux $\hat{\Psi}^{(s)}_{gpd}$ that also consumes GPE during this phase. The second phase is characterised by a change in the diffusive flux direction, which starts to raise the GPE until the granular flow balances into d$\hat{E}^{(s)}_{gp}/\text{d}\hat{t}\rightarrow 0$. This equilibrium is shown in the bottom panel of \cref{fig:energetics}($c$), that illustrates ${\rm d}\hat{E}^{(s)}_{gp}/{\rm d}\hat{t}$ vs time.
\Cref{fig:energetics}($b$) shows time series of the KE production due to shear taken by the small particles $\hat{\Sigma}^{(s)}_{k}$ and the KE dissipation rate owing to friction $\hat{\Psi}^{(s)}_{\mu}$. The magnitude of KE energetics is larger than the GPE energetics, showing that the system is dominated by particle-particle interactions, with a short-lived transient phase, in which shear production is slightly higher than friction dissipation. This results in the KE reaching a shear-friction equilibrium significantly faster than the GPE equilibrium, which is hindered due to the slow pace segregation and the counter effect of diffusive remixing, as shown in \cref{fig:energetics}($c$).

At the equilibrium state, the KE injected by shear is used to mobilise grains horizontally and maintain the bidisperse granular material fully or partially segregated across the sheared layer $h$. Such balanced state\textemdash and therefore its GPE\textemdash should strongly depend on the competition between segregation and diffusion fluxes, which fosters mixing within the granular bulk. This competition is condensed by the segregation-rheology P\'eclet number $Pe_{sr} = Pe/ Sc$.

\begin{figure}
\raggedright
\captionsetup{width=\linewidth}
\SetLabels
\L (0.008*0.99) $(a)$\\
\L (0.008*0.25) \rotatebox{90}{$\rm Energy~partition$}\\
\L (0.68*0.47) \rotatebox{90}{$\mathscr{M}_\phi$}\\
\L (0.17*-0.02) $Pe_{sr}$\\
\L (0.08*0.31) \rotatebox{49}{\color{RawSienna}{$\hat{E}^{(s)}_{k}/\hat{E}^{(s)}_{me}$}}\\
\L (0.1*0.82) \rotatebox{-18}{\color{Blue}{$\hat{E}^{(s)}_{gp}/\hat{E}^{(s)}_{me}$}}\\
\L (0.33*0.99) $(b)$\\
\L (0.327*0.28) \rotatebox{90}{$Ri=\hat{E}^{(s)}_{gp}/\hat{E}^{s}_{k}$}\\
\L (0.23*0.4) {$R_{d}$}\\
\L (0.505*-0.02) $Pe_{sr}$\\
\L (0.76*0.67) $\mu_{0}=$\\
\L (0.68*0.99) $(c)$\\
\L (0.485*0.36) {$R_{d}$}\\
\L (0.84*-0.02) $Pe$\\
\endSetLabels
\strut\AffixLabels{\centerline{\includegraphics[width=0.99\linewidth]{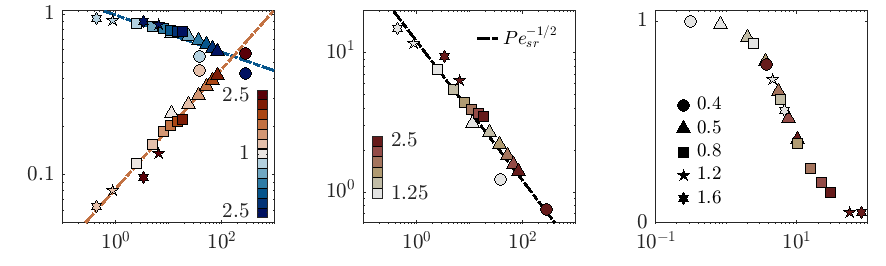}}}
\caption{Results for all 18 numerical solutions. (\emph{a}) Mechanical energy partition vs the segregation-rheology Péclet number $Pe_{sr}$. Reddish markers denote $\hat{E}^{(s)}_{k}/\hat{E}^{(s)}_{me}$, blueish markers denote $\hat{E}^{(s)}_{gp}/\hat{E}^{(s)}_{me}$, with $\hat{E}^{(s)}_{me}=\hat{E}^{(s)}_{k}+\hat{E}^{(s)}_{gp}$. (\emph{b}) Ratio of the mechanical energy components, i.e. the Richardson number $Ri$ vs $Pe_{sr}$. (\emph{c}) Degree of mixing $\mathscr{M}_{\phi}$ vs the Péclet number $Pe$.}
\label{fig:Peclet}
\end{figure}
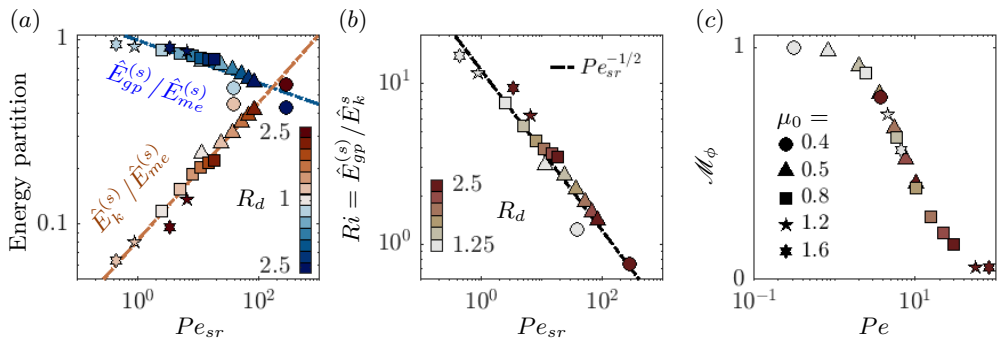

\Cref{fig:Peclet}(\emph{a}) shows the partition of the mechanical energy components of the small particle species versus $Pe_{sr}$. For low-$Pe_{sr}$ ($\leq 10$), the KE is substantially smaller than the GPE, whereas for high-$Pe_{sr}$ ($\geq 10^{2}$), the KE scales with or surpasses the GPE. 
We therefore expect the segregation-mixing state to be sensible to the energy partition and, in particular, to the ratio between GPE and KE, termed as the bulk Richardson number of the granular flow $Ri \equiv \hat{E}^{(s)}_{gp}/ \hat{E}^{(s)}_{k}$. \Cref{fig:Peclet}(\emph{b}) plots $Ri$ vs $Pe_{sr}$ for our numerical solutions. The results exhibit the distinct scaling relationship $Ri \propto Pe_{sr}^{-1/2}$ for three decades of $Pe_{sr}$. From this relationship, the energy partition of the granular flow can be readily predicted since $Pe_{sr}$ is determined from prescribed system parameters. 

Yet, how does the energy partition relate to the segregation-mixing state? We investigate this question through the `degree of mixing'  $\mathscr{M}_{\phi}=1-\left(\sigma_{\phi}/\sigma_{sg}\right)^{2}$, with $\sigma^{2}_{sg}=1/2$ the variance of the fully segregated state of the species concentration $\phi_{s}$ and $\sigma^{2}_{\phi}$ the variance of $\phi_{s}$ at the equilibrium. Thus, $\mathscr{M}_{\phi}=0$ denotes a perfectly segregated state and $\mathscr{M}_{\phi}=1$ a perfectly mixed state. It is apparent that the `degree of segregation' in the system can be quantified from the reciprocal $\mathscr{S}_{\phi}=1-\mathscr{M}_{\phi}$. To focus on the segregation-mixing state attained by the system, we examine $\mathscr{M}_{\phi}$ in terms of the Péclet number $Pe$ instead of $Pe_{sg}$. This allows us to isolate the mechanisms controlling mass distribution and GPE, excluding the friction needed to account for KE. \Cref{fig:Peclet}(\emph{c}) shows $\mathscr{M}_{\phi}$ vs $Pe$. The results show that $Pe$ strongly controls the degree of mixing, characterized by a nonlinear yet monotonic relationship. Within our parameter space, the maximum mixing state is attained at $Pe \simeq 1$, with $\mathscr{M}_{\phi} \approx 0.99$. Conversely, the minimum mixing state, $\mathscr{M}_{\phi}\approx 0.05$, is achieved at $Pe \simeq 10^{2}$, indicating the onset of a saturation state. Statistically, we expect $\mathscr{M}_{\phi}$ not to be zero at high-$Pe$ values as the transition region between small and large particles experiences a persistent remixing-segregation process, as seen in laboratory experiments \citep[e.g.][]{Golick09,ferdowsi2017}.  
\vspace{-0.2cm}

\section{Concluding remarks}
\label{sec:Conclusion}
This paper introduces a continuum framework for the energetics of particle-size segregation in granular flows, providing general analytical expressions for the energetics governing the mechanical energy budget in the system. The framework builds upon the convective-diffusion equation for segregation and a recent segregation scaling law \citep{Trewhela21a}, enabling the study of the complex physics of bidisperse granular flows from a mechanical energy perspective. 

We illustrate the framework's applicability by studying the energetics and mechanical energy partition in shear-driven bidisperse granular flows. Numerical solutions exploring different friction coefficients $\mu_{0}$ and particle-size ratios $R_{d}$ reveal: (i) the existence of distinctive phases in the segregation-mixing energy pathways and (ii) that the ratio between GPE and KE at equilibrium follows the scaling relationship $\hat{E}^{(s)}_{gp} / \hat{E}^{(s)}_{k} \propto (Sc/Pe)^{1/2}$, for $0.4\leq Pe/Sc\leq 300$. Our findings hints that the energy partition of bidisperse granular flows can be predicted from the P\'eclet number $Pe$ and the Schmidt number $Sc$, providing a powerful tool for controlling and understanding segregation-mixing states in granular flows across various engineering applications and geophysical systems.\\

\noindent\textbf{Acknowledgements}. H.N.U.~was~supported~by~the~UPenn~start-up grant. \\

\noindent\textbf{Fundings}. T.T. received support from Agencia Nacional de Investigaci\'on y Desarrollo (ANID) through Fondecyt Initiation Project 11240630.\\


\noindent\textbf{Declaration of interests}. The authors report no conflict of interest.
\vspace{-0.25cm}

\section*{Author ORCIDs}

Tom\'as Trewhela ~~ \href{https://orcid.org/0000-0002-7461-8570}{https://orcid.org/0000-0002-7461-8570}\noindent\\
Hugo N. Ulloa ~~ \href{https://orcid.org/0000-0002-1995-6630}{https://orcid.org/0000-0002-1995-6630}\noindent\\
\vspace{-0.5cm}


\appendix

\bibliographystyle{jfm}
\bibliography{jfm}

\end{document}